\documentclass[oneside,11pt,reqno]{amsart}
\usepackage{amssymb,amsmath,amsthm,bbm,enumerate,mdwlist,url,multirow,hyperref,amsthm}
\usepackage[pdftex]{graphicx}
\usepackage[shortlabels]{enumitem}

\theoremstyle{definition}
\newtheorem{definition}{Definition}%Extra square-bracket argument achives that the numbering is the same as for definition (single uniform counter). 
\theoremstyle{theorem}
\newtheorem{proposition}[definition]{Proposition}

\newtheorem{theorem}[definition]{Theorem}

\numberwithin{equation}{section}
\theoremstyle{remark}
\newtheorem{remark}[definition]{Remark}
\newtheorem{question}[definition]{Question}
\def\PP{\mathsf P}
\def\Ruin{\mathrm{Ruin}}
\def\Exp{\mathrm{Exp}}
\def\var{\mathrm{var}}
\def\cov{\mathrm{cov}}
\def\EE{\mathsf E}
\def\LL{\mathcal L}
\def\FF{\mathcal F}
\begin{document}
\title{Ruin under stochastic dependence between premium and claim arrivals}

\author{Matija Vidmar}
\address{Department of Mathematics, University of Ljubljana, Slovenia}
\address{Institute for Mathematics, Physics and Mechanics, Ljubljana, Slovenia}
\email{matija.vidmar@fmf.uni-lj.si}

\begin{abstract}
We investigate, focusing on the ruin probability, an adaptation of the Cram\'er-Lundberg model for the surplus process of an insurance company, in which, \emph{conditionally} on their intensities, the two mixed Poisson processes governing the arrival times of the premiums and of the claims respectively, are independent. Such a model exhibits a stochastic dependence between the aggregate premium and claim amount processes. An explicit expression for the ruin probability is obtained when the claim and premium sizes are exponentially distributed.
\end{abstract}



\maketitle

\section{Introduction}
A classical model for the capital surplus of an insurance company is the Cram\'er-Lundberg process -- the sum of an initial capital, of a deterministic positive premium drift and of a compound Poisson process of negative-valued claims. Various generalizations of this benchmark model exist: replacing the homogeneous Poisson process governing the claim arrivals by a more general counting process, adding a drift-diffusion component to reflect capital gains/losses, randomizing the premium income, to name just a few. A fundamental quantity is the ruin probability. \cite{andreas,asmussen,dickson,rolski} In this short note we investigate a particular model that features stochastic dependence of the premium and claim arrival processes, and whose ruin probability, in the special case of exponentially distributed claims and stochastic premiums, allows for an explicit solution -- see Theorem~\ref{theorem}. We find that the determination of said ruin probability is intimately related to the study of the same in the classical Cram\'er-Lundberg model, in which, however, the claims may be negative, and the premium drift may be zero -- Proposition~\ref{proposition}. 

As regards the relevant literature, publications concerning  models for the surplus process in which the premium income is stochastic, but \emph{independent} of the claim amount process, include \cite{boikov,temnov,bondarev,labbe,albrechergerber,bao}. On the other hand, models featuring \emph{stochastic dependence} between premium income and the claim amount process have been studied in \cite{zhang}, where the premium sizes and interclaim times are controlled by the claim sizes; and in \cite{zhou}, where the current premium rate is adjusted after a claim occurs, the adjusted rate being determined by the size of the claim. See also the references in these cited works. 

Our contribution is the analysis of another kind of stochastic dependence that can occur between the aggregate premium and claim amount processes (namely, one that can exist between the premium and claim arrival times). For the counting processes determining the arrivals of the premiums and claims we choose mixed Poisson processes, a generalization of the class of homogeneous Poisson processes that is commonly used (amongst others) in the insurance context. This special probabilistic structure of our model, while a loss on generality, allows for (comparatively speaking) more tractable results. 

We have mentioned the investigation of our model is related to the classical Cram\'er-Lundberg one but with two-sided jumps. Our contribution in this regard is a fully general (assuming only a nonnegative drift and integrability of the claim and premium sizes) integral (renewal) equation for the ruin probability. Despite the prevalence  \cite[Paragraph~XII.4b]{asmussen} \cite{perry,zhang-yang,zacks,yin} of the two-sided compound Poisson process as a component in the model of the surplus process, this result appears not yet to have been recorded  for our form of the surplus process and under the indicated generality (specifically, we have no continuity assumptions on the distribution of the claim and premium sizes). Its special case of the positive jumps and the negative jumps being both exponentially distributed, however, is very well-known and has been studied, in far greater generality than just that which concerns the ruin probability, see e.g. \cite[Section~4]{yin} \cite{kou} and the reference therein. For the sake of completeness, rather than just citing the relevant formula for the ruin probability of such a double-exponential model (that we ultimately need to obtain Theorem~\ref{theorem}), we instead briefly solve the mentioned integral-renewal equation for this case. 

\section{The model \& notation}\label{section:model}
We fix a probability space $(\Omega,\FF,\PP)$ and thereon the following independent random elements: 
\begin{itemize}
\item an iid sequence $Z=(Z_i)_{i\in \mathbb{N}}$ of random variables with values in $(0,\infty)$, of finite mean;
\item another iid sequence $Y=(Y_i)_{i\in \mathbb{N}}$ of random variables with values in $(0,\infty)$, of finite mean;
\item a quadruplet $((M=(M_t)_{t\in[0,\infty)},L=(L_t)_{t\in [0,\infty)},\Delta,\Gamma)$, where $\Delta$ and $\Gamma$ have values in $(0,\infty)$, whilst $M$ and $L$ are, conditionally on $(\Delta,\Gamma)$, independent homogeneous Poisson processes with intensities $\Delta$ and $\Gamma$, respectively. 
\end{itemize}
Note that we may realize $(M,L)$ (assuming $(Z,Y,\Delta,\Gamma)$ given) by taking, independently of $(Z,Y,\Delta,\Gamma)$, two independent Poisson processes $P^1$ and $P^2$ of unit intensity, and then setting $M_t=P^1_{t\Delta}$ and $L_t=P^2_{t\Gamma }$ for $t\in [0,\infty)$.

The surplus process $K=(K_t)_{t\in [0,\infty)}$ is then defined, for a given initial capital $u\in [0,\infty)$ and fixed deterministic premium rate $c\in [0,\infty)$, as follows: $$K_t:=u+ct+\sum_{i=1}^{M_t}Y_i-\sum_{i=1}^{L_t}Z_i,\quad t\in [0,\infty).$$ In this way the capital inflow has both a deterministic (linear drift) as well as a stochastic component. 

Finally we let $\Ruin:=\{\inf_{t\in [0,\infty)}K_t<0\}$ be the event of ruin and then define $\psi(u):=\PP(\Ruin)$ and $\phi(u):=1-\psi(u)$, the ruin and non-ruin probabilities respectively. It makes sense to further stipulate $\psi(v):=1$ and then $\phi(v):=0$ for $v\in (-\infty,0)$. 

In terms of general notation: $\PP_T$ and $F_T$ will, respectively, denote the law and distribution function of a random variable $T$.

\section{Results}

\begin{remark}
Wald's identity implies that $\EE K_t=u+t(c+\EE Y_1\EE \Delta-\EE Z_1 \EE\Gamma)$. Assuming $Y_1$, $Z_1$, $\Lambda$ and $\Gamma$ are all square-integrable, the relation $\var(K_t)=\EE [ \var(K_t\vert(\Delta,\Gamma))]+\var(\EE[K_t\vert (\Delta,\Gamma)])$ and expressions for the mean and variance of compound Poisson sums further yield $\var(K_t)=\EE [Y_1^2](\EE[\Delta]t+t^2\var(\Delta))+\EE [Z_1^2](\EE[\Gamma]t+t^2\var(\Gamma))-2t^2\EE Y_1\EE Z_1\cov(\Delta,\Gamma)$.
\end{remark}
Let us now focus on the ruin probability. By conditioning on $(\Delta,\Gamma)$, exploiting all the relevant independences and taking into account properties of conditional expectations, we find that $$\phi(u)=\EE[\PP[\Omega\backslash \Ruin\vert(\Delta,\Gamma)]]=\EE[\phi^{\Gamma,\Delta}(u)]$$ where for $\{\gamma,\delta\}\subset (0,\infty)$, $\phi^{\gamma,\delta}(u)$ is our $\phi(u)$ in the special case when $\Gamma$ is identically equal to $\gamma$ and $\Delta$ is identically equal to $\delta$. Now, thanks to standard results concerning marked Poisson processes, $\phi^{\gamma,\delta}(u)$ may be seen as the probability of non-ruin in the model in which the surplus process is given by $$C_t:=u+ct-\sum_{i=1}^{N_t}X_i\text{ for }t\in [0,\infty),$$ where $X=(X_i)_{i\in \mathbb{N}}$ is an iid sequence of random variables with common law $\frac{\gamma}{\gamma+\delta}\PP_{Z_1}+\frac{\delta}{\gamma+\delta}\PP_{-Y_1}$, independent of the homogeneous Poisson process $N$ of intensity $\lambda:=\delta+\gamma$. This is a classical Cram\'er-Lundberg model, except for the fact that the claim sizes can be negative and the premium rate can be zero.

For $n\in \mathbb{N}$ let $Q_n:=X_n-cW_n$, $S_n:=\sum_{k=1}^nQ_k$, where $W_n$ is the $n$-th inter-arrival time of $N$. Clearly, thanks to $c\geq 0$, $\phi^{\gamma,\delta}(u)$ is equal to $$\PP(S_n\leq u\text{ for all }n\in\mathbb{N})=\PP(Q_1\leq u,S_n-Q_1\leq u-Q_1\text{ for all }n\in\mathbb{N}_{\geq 2})$$ and by  conditioning on $(X_1,W_1)$ exploiting the fact that $((X_i,W_i))_{i\in \mathbb{N}}$ is an iid sequence $$=\int_0^\infty dw\lambda e^{-\lambda w}\int_{(-\infty,u+cw]}dF_{X_1}(x)\phi^{\gamma,\delta}(u+cw-x)$$ and further (assuming, for the time being, $c>0$ and effecting the change of variables $z=u+cw$)
$$=\frac{\lambda}{c}e^{u\lambda/c}\int_u^\infty dze^{-\lambda z/c}\int_{(-\infty,z]}dF_{X_1}(x)\phi^{\gamma,\delta}(z-x).$$ By an integration by parts for the continuous (locally of) finite variation functions $u\mapsto e^{u\lambda/c}=\frac{\lambda}{c}\int_{-\infty}^ue^{z\lambda/c}dz$ and $u \mapsto\int_u^\infty dze^{-\lambda z/c}\int_{(-\infty,z]}dF_{X_1}(x)\phi^{\gamma,\delta}(z-x)$, mapping $[0,\infty)\to \mathbb{R}$, we may rewrite this into $$\phi^{\gamma,\delta}(u)=\phi^{\gamma,\delta}(0)+\frac{\lambda}{c}\int_0^u\phi^{\gamma,\delta}(z)dz-\frac{\lambda}{c}\int_0^udz\int_{(-\infty,z]}dF_{X_1}(x)\phi^{\gamma,\delta}(z-x),$$ where we have also used associativity of integration. Then by Tonelli's theorem and a subsequent change of variables in the Lebesgue integrals \footnotesize $$\phi^{\gamma,\delta}(u)-\phi^{\gamma,\delta}(0)-\frac{\lambda}{c}\int_0^u\phi^{\gamma,\delta}(z)dz%=-\frac{\lambda}{c}\left[\int_{(-\infty,u]}dF_{X_1}(x)\int_{x\lor 0}^udz\phi^{\gamma,\delta}(z-x)\right]$$ $$
=-\frac{\lambda}{c}\left[\int_{(-\infty,0)}dF_{X_1}(x)\int_{-x}^{u-x}dz\phi^{\gamma,\delta}(z)+\int_{[0,u]}dF_{X_1}(x)\int_{0}^{u-x}dz\phi^{\gamma,\delta}(z)\right]$$\normalsize so that another application of Tonelli's theorem yields \footnotesize $$\phi^{\gamma,\delta}(u)-\phi^{\gamma,\delta}(0)-\frac{\lambda}{c}\int_0^u\phi^{\gamma,\delta}(z)dz$$ $$=-\frac{\lambda}{c}\left[\int_0^udz\phi^{\gamma,\delta}(z)\frac{\delta}{\gamma+\delta}F_{Y_1}(z)+\int_u^\infty dz \phi^{\gamma,\delta}(z)\frac{\delta}{\gamma+\delta}(F_{Y_1}(z)-F_{Y_1}(z-u))+\int_0^udz\frac{\gamma}{\gamma+\delta}\phi^{\gamma,\delta}(z)F_{Z_1}(u-z)\right],$$\normalsize i.e. \footnotesize $$\phi^{\gamma,\delta}(u)-\phi^{\gamma,\delta}(0)-\frac{\lambda}{c}\int_0^u\phi^{\gamma,\delta}(z)dz$$ $$=-\frac{\lambda}{c}\left[\int_0^\infty dz\phi^{\gamma,\delta}(z)\frac{\delta}{\gamma+\delta}F_{Y_1}(z)-\int_u^\infty dz \phi^{\gamma,\delta}(z)\frac{\delta}{\gamma+\delta}F_{Y_1}(z-u)+\int_0^udz\frac{\gamma}{\gamma+\delta}\phi^{\gamma,\delta}(z)F_{Z_1}(u-z)\right]$$ 
$$=-\frac{\lambda}{c}\left[\int_0^\infty dz(\phi^{\gamma,\delta}(z)-\phi^{\gamma,\delta}(z+u))\frac{\delta}{\gamma+\delta}F_{Y_1}(z)+\int_0^udz\frac{\gamma}{\gamma+\delta}\phi^{\gamma,\delta}(u-z)F_{Z_1}(
z)\right].$$ \normalsize In other words %\footnotesize $$c(\phi^{\gamma,\delta}(u)-\phi^{\gamma,\delta}(0))-\delta\int_0^u\phi^{\gamma,\delta}(u-z)dz=\delta\int_0^\infty dz(\phi^{\gamma,\delta}(z+u)-\phi^{\gamma,\delta}(z))F_{Y_1}(z)+\gamma\int_0^udz\phi^{\gamma,\delta}(u-z)\overline{F}_{Z_1}(z)$$ or
 $$c(\phi^{\gamma,\delta}(u)-\phi^{\gamma,\delta}(0))=-\delta\int_0^\infty dz(\phi^{\gamma,\delta}(z+u)-\phi^{\gamma,\delta}(z))\overline{F}_{Y_1}(z)+\gamma\int_0^udz\phi^{\gamma,\delta}(u-z)\overline{F}_{Z_1}(
z).$$ \normalsize

From the theory of random walks (whether or not $c>0$): if $c+\delta \EE Y_1\leq \gamma \EE Z_1$, then $\phi^{\gamma,\delta}(u)=0$; otherwise $\lim_{u\to \infty}\phi^{\gamma,\delta}(u)=1$. Assume the latter. Letting $u\uparrow \infty$ in the last display, plugging it back in, and expressing everything in terms of $\psi^{\gamma,\delta}:=1-\phi^{\gamma,\delta}$, we obtain $$c\psi^{\gamma,\delta}(u)=-\delta\int_0^\infty dz\psi^{\gamma,\delta}(z+u)\overline{F}_{Y_1}(z)+\gamma\int_0^udz \psi^{\gamma,\delta}(u-z)\overline{F}_{Z_1}(
z)+\gamma\int_u^\infty dz\overline{F}_{Z_1}(z),$$ which may also be rewritten as  %$$c\psi^{\gamma,\delta}(u)=\int_{-\infty}^u\psi^{\gamma,\delta}(u-z)(\mathbbm{1}_{(0,\infty)}(z) \gamma\overline{F}_{Z_1}(z)-\mathbbm{1}_{(-\infty,0)}(z)\delta\overline{F}_{Y_1}(-z))dz+\gamma\int_u^\infty dz\overline{F}_{Z_1}(z)$$ which is the same as 
(recall that $\psi^{\gamma,\delta}(v)=1$ for $v\in (-\infty,0)$)
\begin{equation}\label{equation:integral}
c\psi^{\gamma,\delta}(u)=\int_{\mathbb{R}}\psi^{\gamma,\delta}(u-z)dG(z),
\end{equation}
where for $z\in \mathbb{R}$ we have introduced $G(z):=\int_0^z\gamma\overline{F}_{Z_1}(y)dy\mathbbm{1}_{(0,\infty)}(z)+\int_0^{-z}\delta\overline{F}_{Y_1}(y)dy\mathbbm{1}_{(-\infty,0)}(z)$. Using dominated convergence we may pass to the limit $c\downarrow 0$ and conclude \eqref{equation:integral} remains valid in the case when $c=0$, provided of course $\delta \EE Y_1>\gamma\EE Z_1$ -- the provisional assumption $c>0$ being now dropped. 

It is worth recording this partial result separately.

\begin{proposition}\label{proposition}
Let the surplus process $C$ be given by $$C_t:=u+ct-\sum_{i=1}^{N_t}X_i\text{ for }t\in [0,\infty),$$ where $\{u,c\}\subset [0,\infty)$ and $X=(X_i)_{i\in \mathbb{N}}$ is an iid sequence of random variables with common law $\LL$, independent of the homogeneous Poisson process $N$ of intensity $\lambda\in (0,\infty)$. Let (fixing everything apart from $u$) $\nu(u):=\PP(\inf_{t\in [0,\infty)}C_t<0)$ be the probability of ruin. Assume $\LL(\{0\})=0$, $\{\int_{(0,\infty)} x \LL(dx),-\int_{(-\infty,0)} x\LL(dx)\}\subset (0,\infty)$. If $c\leq \lambda \int x\LL(dx)$, then $\psi(u)=1$, otherwise 
\begin{equation}\label{eq:renewal}
c\nu(u)=\int_{\mathbb{R}}\nu(u-z)dG(z)
\end{equation}
where for $z\in \mathbb{R}$, $G(z):=\lambda(\int_0^z\LL(y,\infty)dy\mathbbm{1}_{(0,\infty)}(z)+\int_0^{-z}\LL(-\infty,-y)dy\mathbbm{1}_{(-\infty,0)}(z))$, and we understand $\nu(v)=1$ for $v\in (-\infty,0)$.
\end{proposition}
\begin{remark}
If there exists $r\in (0,\infty)$ satisfying $\int e^{yr}\LL(dy)\frac{\lambda}{\lambda+cr}=1$, equivalently $c=\int e^{rz}dG(z)$, i.e. if there exists an adjustment coefficient, it follows by the usual argument that one has the Lundberg bound $\psi(u)\leq e^{-ru}$. In this case, if furthermore $c>0$, we may define the exponentially tilted function $H(z):=\frac{\lambda}{c}\left(\int_0^ze^{ry}\LL(y,\infty)dy\mathbbm{1}_{(0,\infty)}(z)+\int_0^{-z}e^{-ry}\LL(-\infty,-y)dy\mathbbm{1}_{(-\infty,0)}(z)\right)$ for $z\in \mathbb{R}$, such that $dH$ becomes a signed measure of mass $dH(\mathbb{R})=1$, and \eqref{eq:renewal} may be rewritten as a kind-of renewal equation
\begin{equation}\label{kind-of-renewal} 
\xi=\xi \star H\text{ on }[0,\infty)
\end{equation}
for the bounded function $\xi$ defined by $\xi(v):=e^{rv}\nu(v)$ for $v\in \mathbb{R}$. 
\end{remark}
Let us now particularize to the case when, with  $\{a,b\}\subset (0,\infty)$, $Y_1\sim \Exp(a)$ and $Z_1\sim \Exp(b)$. We may rewrite \eqref{equation:integral} into 
\begin{equation}\label{on-path-to-de:one}
c\psi^{\gamma,\delta}(u)=-\delta e^{au}\int_u^\infty\psi^{\gamma,\delta}(z)e^{-az}dz+\gamma e^{-bu}\int_0^u\psi^{\gamma,\delta}(z)e^{bz}dz+\frac{\gamma}{b}e^{-bu}.
\end{equation}
Since $\psi^{\gamma,\delta}$ is bounded, it follows from the right hand-side of \eqref{on-path-to-de:one}, by dominated convergence, that $\psi^{\gamma,\delta}$ is even continuous, and then again from the right hand-side of \eqref{on-path-to-de:one}, by the fundamental theorem of calculus, that $\psi^{\gamma,\delta}$ is differentiable (on $[0,\infty)$; we mean of course the right derivative at $0$). Differentiating \eqref{on-path-to-de:one} and then adding back to the  identity thus obtained the identity \eqref{on-path-to-de:one} multiplied by $b$, we find (where we omit certain evaluations at $u$ for brevity of notation)
\begin{equation}\label{on-path-to-de:two}
bc\psi^{\gamma,\delta}+c(\psi^{\gamma,\delta})'=(\delta+\gamma)\psi^{\gamma,\delta}-(a+b)\delta e^{au}\int_u^\infty\psi^{\gamma,\delta}(z)e^{-az}dz.
\end{equation}
Repeating, mutatis mutandis, this exercise one more time, we find that $\psi^{\gamma,\delta}$ is of class $C^2$ on $[0,\infty)$ and satisfies 
\begin{equation}\label{DE}
c(\psi^{\gamma,\delta})''+(c(b-a)-\delta-\gamma)(\psi^{\gamma,\delta})'+(-b\delta+a\gamma-cba)\psi^{\gamma,\delta}=0
\end{equation}
thereon. Now, since the net profit condition $c+\delta/a>\gamma/b$ is being assumed, it is trivial to check that the equation for the adjustment coefficient, $c+\frac{\delta}{a+r}=\frac{\gamma}{b-r}$ in $r\in \mathbb{R}\backslash \{-a,b\}$, has precisely two roots, one in $(0,b)$, henceforth denoted $r^{\gamma,\delta}$, and the other in $(-\infty,-a)$, the two moreover, modulo a sign change, being also precisely the two zeros of the characteristic polynomial of \eqref{DE}. It follows from the general theory of homogeneous second order linear differential equations with constant coefficients and the boundedness of $\psi^{\gamma,\delta}$ that $\psi^{\gamma,\delta}(u)=Ae^{-r^{\gamma,\delta}u}$ for some $A\in [0,\infty)$. Plugging this into \eqref{on-path-to-de:one}, we find that $A=1-r^{\gamma,\delta}/b$.

Bringing everything together, we arrive at

\begin{theorem}\label{theorem}
The ruin probability in the model of Section~\ref{section:model}, in which, with  $\{a,b\}\subset (0,\infty)$, $Y_1\sim \Exp(a)$ and $Z_1\sim \Exp(b)$, is given by the expression $$\psi(u)=\int_{(0,\infty)^2}\left[\left(1-\frac{r^{\gamma,\delta}}{b}\right)e^{- r^{\gamma,\delta}u}\mathbbm{1}_{(-\infty,c)}\left(\frac{\gamma}{b}-\frac{\delta}{a}\right)\right]\PP_{(\Gamma,\Delta)}(d\gamma,d\delta)+\PP\left(\frac{\Gamma}{b}-\frac{\Delta}{a}\geq c\right),$$ where $$r^{\gamma,\delta}=
\frac{c(b-a)-\delta-\gamma+\sqrt{((a+b)c+\delta)^2-2((a+b)c-\delta)\gamma+\gamma^2}}{2c},$$ assuming $c>0$, and by 
$$\psi(u)=\int_{(0,\infty)^2}\left[\frac{1+a/b}{1+\delta/\gamma}e^{-\frac{b\delta-a\gamma}{\delta+\gamma}u}\mathbbm{1}_{(-\infty,0)}\left(\frac{\gamma}{b}-\frac{\delta}{a}\right)\right]\PP_{(\Gamma,\Delta)}(d\gamma,d\delta)+\PP\left(\frac{\Gamma}{b}-\frac{\Delta}{a}\geq 0\right),$$  when $c=0$.
\end{theorem}

\begin{question}
Relation \eqref{eq:renewal} does not extend to $u\in (-\infty,0)$ (as one can e.g. check explicitly in the case of the double-sided exponential distribution). Nevertheless \eqref{kind-of-renewal} does not seem too far-off from a classical renewal equation  on the whole real line. What could be said, in general, about the asymptotics of $\xi$ based on this equation?
\end{question}
\bibliographystyle{plain}
\bibliography{Biblio_ruin}
\end{document}